\documentclass[prl]{revtex4}

\usepackage{graphicx}
\usepackage{dcolumn}
\usepackage{bm}

\begin{document}
\renewcommand{\theequation}{\arabic{equation}}

\title{Radiative Processes in External Gravitational Fields\\}

\author{Giorgio Papini}
\altaffiliation[Electronic address:]{papini@uregina.ca}
\affiliation{Department of Physics, University of Regina, Regina,
Sask, S4S 0A2, Canada} \affiliation{Prairie Particle Physics
Institute, University of Regina, Regina, Sask, S4S 0A2, Canada}
\affiliation{International Institute for Advanced Scientific
Studies, 89019 Vietri sul Mare (SA),
 Italy.}

\date{\today}

\begin{abstract}
Kinematically forbidden processes may be
allowed in the presence of external gravitational fields.
These can be taken into account by introducing
generalized particle momenta.
The corresponding transition probabilities can then be calculated to all
orders in the metric deviation from the field-free expressions by simply replacing the particle momenta
with their generalized counterparts.
The procedure applies to particles of any spin and to
any gravitational fields. Transition probabilities, emission power and spectra are, to leading
order, linear in the metric deviation. It is also shown how a small dissipation term in the particle wave equations
can trigger a strong back-reaction that introduces resonances in the radiative process and deeply affects the resulting gravitational background.

\end{abstract}

 \maketitle

\begin{figure}
\centering
\includegraphics[width=0.3\textwidth]{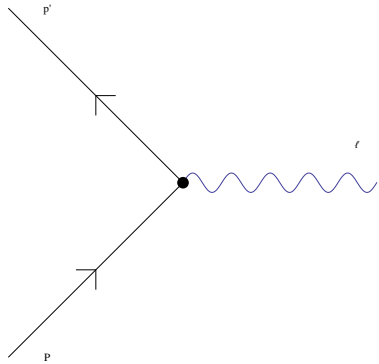}
\caption{\label{fig:Feynman2} $p'$ and $\ell$ are the outgoing fermion and
photon and $ P$ indicates the incoming fermion.}
\end{figure}

\setcounter{equation}{0}
Processes in which massive, on-shell particles emit a photon
according to Fig.\ref{fig:Feynman2} are examples of kinematically forbidden transitions
that remain so
unless the dispersion relations of at least one of the particles involved are altered.
This possibility presents itself when particles travel in a medium or in
an external gravitational field.

The action of gravitational fields on a particle's dispersion relations
can be studied by solving the respective covariant wave equation.
This can be done exactly to first order in the metric deviation $ \gamma_{\mu\nu}= g_{\mu\nu}
- \eta_{\mu\nu}$, where $ \eta_{\mu\nu}$ is the Minkowski metric \cite{caipap,dinesh,papa,papb,pap1,pap2,pap3,pap4}. The solutions
contain the gravitational field in a phase operator that alters in effect a particle four-momentum
by acting on
the wave function of the field-free equations. This result applies equally well to fermions and bosons and can be extended to all orders in $\gamma_{\mu\nu}$.
The calculation
of even the most elementary Feynman diagrams does therefore require an appropriate treatment when gravitational fields are present. While the inclusion of external electromagnetic fields
has met with success in the case of static (Coulomb) fields \cite{jauch},
and can be easily carried out for static (Newtonian) fields, no systematic attempts have been made for relativistic gravity.
The procedure developed below is intended to fill in part this gap and applies to weak, static and non-static gravitational fields.

Let us assume, for simplicity, that $P$ in Fig.\ref{fig:Feynman2}
is an incoming fermion and that the photon $\ell$ and outgoing
fermion $p'$ are produced on-shell. The solution of the covariant
Dirac equation, exact to ${\cal O}(\gamma_{\mu\nu})$, is \cite{pap1}
\begin{equation}\label{Psi}
  \Psi(x)=-\frac{1}{2m}\left(-i\gamma^\mu(x){\cal
  D}_\mu-m\right)e^{-i\Phi_T}\Psi_0(x)\equiv \hat{T}\Psi_{0}\,,
\end{equation}
where ${\cal D}_\mu=\nabla_\mu+i\Gamma_\mu (x)$, $\nabla_\mu$ is
the covariant derivative, $\Gamma_{\mu}(x)$ the spin connection
and the matrices $\gamma^{\mu}(x)$ satisfy the relations
$\{\gamma^\mu(x), \gamma^\nu(x)\}=2g^{\mu\nu}(x)$. Both
$\Gamma_\mu(x)$ and $\gamma^\mu(x)$ can be obtained from the usual
constant Dirac matrices by using the vierbein fields $e_{\hat
\alpha}^\mu$ and the relations
\begin{equation}\label{II.2}
\gamma^\mu(x)=e^\mu_{\hat \alpha}(x) \gamma^{\hat \alpha}=\delta^{\mu}_{\hat{\alpha}}+h^{\mu}_{\hat{\alpha}}(x)\,,\qquad
\Gamma_\mu(x)=-\frac{1}{4} \sigma^{{\hat \alpha}{\hat \beta}}
e^\nu_{\hat \alpha}e_{\nu\hat{\beta};\, \mu}\,,
\end{equation}
where $\sigma^{{\hat \alpha}{\hat \beta}}=\frac{i}{2}[\gamma^{\hat
\alpha}, \gamma^{\hat \beta}]$. A semicolon and a comma are
also used as alternative ways to indicate covariant and
partial derivatives respectively. We use units $ \hbar = c = 1$,
the signature of $\eta_{\mu\nu}$ is
$-2\,,\Phi_T=\Phi_s+\Phi_G$,
\begin{equation}\label{PhiG}
 \Phi_s(x)={\cal P}\int_P^x dz^\lambda
\Gamma_\lambda (z)\,\,\,, \Phi_{G}=-\frac{1}{4}\int_P^xdz^\lambda\left[\gamma_{\alpha\lambda,
  \beta}(z)-\gamma_{\beta\lambda, \alpha}(z)\right]L^{\alpha\beta}(z)+
  \frac{1}{2}\int_P^x dz^\lambda\gamma_{\alpha\lambda}k^\alpha\,,
\end{equation}
$ L^{\alpha\beta}(z)=(x^\alpha-z^\alpha)k^\beta-
 (x^\beta-z^\beta)k^\alpha$
and $\Psi_0(x)$ satisfies the usual, flat spacetime Dirac equation.

It is convenient to re-write (\ref{Psi}) in the form $ \Psi(x) =
g(x) exp(-ipx)u_{0}(\vec{p})$, where
\begin{equation} \label{g}
g(x)=\frac{1}{2m}\left[\gamma^{\mu}\left(p_{\mu}+h^{\hat{\alpha}}_{\mu}(x)p_{\hat{\alpha}}+\Phi_{G,\mu}(x)\right)+m\right]e^{-i\Phi_{T}}\,,
\end{equation}
and
\begin{equation}\label{BPhiGDer}
  \Phi_{G, \mu}=-\frac{1}{2}\int_P^x dz^\lambda (\gamma_{\mu\lambda,
  \beta}-\gamma_{\beta\lambda,
  \mu})p^\beta+\frac{1}{2}\gamma_{\alpha\mu}p^\alpha\,.
\end{equation}
It also is convenient to focus on the simple process of Fig.\ref{fig:Feynman2}. We claim that the transition amplitude can
be calculated by introducing the generalized four-momentum
\begin{equation}\label{mom}
P_{\mu}= p_{\mu}+\tilde{h}^{\hat{\alpha}}_\mu
p_{\hat{\alpha}}+\tilde{\Phi}_{G,\mu}-\frac{i}{2}(\tilde{\Phi}_{G}-\tilde{\Phi}^{*}_{G})p_{\mu}\equiv p_{\mu}+\tilde{P}_{\mu}\,,
\end{equation}
for the incoming fermion, as (\ref{g}) itself suggests. The part
that contains the gravitational field is indicated by $\tilde{P}_\mu$. In (\ref{mom}), $
\tilde{h}^{\hat{\alpha}}_\mu$, $\tilde{\Phi}_{G,\mu}$ and
$\tilde{\Phi}_G$ are quantities that must be calculated, once the metric is known. They are related to the Fourier transforms of the
corresponding expressions that appear in (\ref{II.2}), (\ref{PhiG}) and
(\ref{BPhiGDer}). $P_{\mu}$
is not on-shell. In fact
\begin{equation}\label{mass}
P^{\mu}P_{\mu} \equiv m^{2}_e =m^2 +2\left[p^\mu
h_{\mu}^{\hat{\alpha}}p_{\alpha}+ p^\mu \Phi_{G,\mu}-\frac{i m^2}{2}
(\Phi_{G}-\Phi^{*}_{G})\right]\,,
\end{equation}
where $p_{\mu}p^{\mu}=m^2$ because $p_{\mu}$ is the momentum
of the free fermion represented by $ \Psi_{0}(x)$ in (\ref{Psi}).
The transition amplitude is then
\begin{equation}\label{M1}
M_{1} =
-iZe\eta^{\mu\nu}\bar{u}_{0}(\vec{p'})\varepsilon_{\hat{\mu}(\lambda)}\gamma_{\hat{\nu}}g(|\vec{q}|)u_{0}(\vec{p})\,,
\end{equation}
where $\vec{q}\equiv \vec{p}-\vec{p'}-\vec{\ell}$, $\,\varepsilon^{\hat{\mu}}_{(\lambda)}$ represents the
polarization of the photon, and $Ze$ is the charge of the fermion.
A transition amplitude $ M_2 $ must be added to $M_1$ to account for the fact that the contraction in (\ref{M1}) is in general accomplished by means of $g^{\mu\nu}$.
It has been repeatedly calculated in the literature
and is given by \cite{papvall}
\begin{equation}\label{M2}
M_{2}=-iZe\gamma^{\mu\nu}(|\vec{q}|)\bar{u}_{0}(\vec{p'})\varepsilon_{\hat{\mu}(\lambda)}\gamma_{\hat{\nu}}u_{0}(\vec{p})\,.
\end{equation}
$M_{1}$ contains the part $p_{\mu}$ of (\ref{mom}), that comes from $\Psi_{0}$, and a new part that contains the gravitational
contribution due to the propagation of the fermion in the field of the source. The total transition amplitude is given by $M=M_1+M_2$.

The calculation now requires that a metric be selected.

Let us consider the particular instance of a fermion that is
propagating with momentum $p^3 \equiv p$, impact parameter $b\geq
R$ and $x_{2}=0$, from $x_3 =-\infty$ toward a gravitational
source of mass $M$ and radius $R$ placed at the origin and
described by the metric $\gamma_{00}=2\phi\,,
\gamma_{ij}=2\phi\delta_{ij} $, where $\phi=-\frac{GM}{r}$. This
metric is frequently used in lensing problems
\cite{lensing},\cite{pap4}. One finds $ \Gamma_{0}=-1/2 \phi_{,j}
\sigma^{0j}\,,\Gamma_{i}=-1/2 \phi_{,j} \sigma^{ij}$ and $
e^0_{\hat{i}}=0\,,
 e^0_{\hat{0}}=1-\phi\,,
 e^l_{\hat{k}}=\left(1+\phi\right)\delta^l_k $.
All spin matrices are now expressed in terms of ordinary, constant
Dirac matrices. We also assume that the on-shell conditions $
p'_{\mu}p'^{\mu}=m^2\,,\ell_{\mu}\ell^{\mu}=0$ remain valid.
Extension of the calculation to include different particles, or
higher order gravitational contributions to $p'\,,\ell$, and
(\ref{Psi}) can be derived to all orders in $\gamma_{\mu\nu}$.

The Fourier transforms of the quantities that appear in (\ref{g}) must now be
calculated. We obtain
\begin{equation}\label{h}
h^{\hat{\alpha}}_{0}(q)p_{\alpha}=8\pi^2\delta(q_0)\delta(q_x)\delta(q_y)p_0 GM K_0(bq_z)\,, h^{\hat{\alpha}}_{3}(q)p_{\alpha} =8\pi^2\delta(q_0)\delta(q_x)\delta(q_y)p GM K_0(bq_z)\,,
\end{equation}
\begin{equation}\label{phgmu}
\Phi_{G,1}(q)+\Phi^{*}_{G,1}(q)=0\,,\Phi_{G,2}(q)=0\,,\Phi_{G,3}(q)= -8\pi^2\delta(q_0)\delta(q_x)\delta(q_y)\left(\frac{p^2_0}{p}+p\right)GM K_0(bq_z)\,,
\end{equation}
and
\begin{equation}\label{phitr}
 \Phi_{G}(q)=-i16\pi^{2}\delta(q_0)\delta(q_x)\delta(q_y)\left(\frac{2p^{2}_0}{p}-p_0+p\right)\frac{GM}{bq_z}K_0(bq_z)\,.
 \end{equation}
Four-momentum conservation to zeroth order only is required because (\ref{h}), (\ref{phgmu}) and (\ref{phitr}) are already of ${\cal O}(\gamma_{\mu\nu})$. We
further approximate the Bessel function $K_0(bq_z)\simeq \sqrt{\pi/2bq_z}e^{-bq_z} [1-1/8bq_z +...]$, itself a distribution, by $K_0(bq_z)\simeq \delta(bq_z)$ and eliminate $\delta^4(q)$
from (\ref{h}), (\ref{phgmu}) and (\ref{phitr}). Conservation of energy-momentum will reappear as a factor $(2\pi)^4 \delta^{4}(q)$ in the expression for the radiated power $W$ defined below.
The term $\Phi_{G}(q)$ does however diverge even more rapidly than the other Fourier transforms for small momentum transfers and will be dropped in what follows.
 This behavior is well known and is related to the infinite range of the Newtonian potential and the use of a plane wave for $\Psi_{0}(x)$.
By removing $\delta^{4}(q)$ from $h^{\hat{\alpha}}_{\mu}$ and $\Phi_{G,\mu}(q)$ we obtain  $ \tilde{h}^{\hat{\alpha}}_\mu$ and $\tilde{\Phi}_{G,\mu}$ of (\ref{mom}).
We find
\begin{equation}\label{mom1}
P_0 \simeq p_0 +4\pi \frac{GM}{b}p_0=p_0+\tilde{P}_0\,,
P_1\simeq 4\pi\frac{GM}{b}\left(\frac{p_0^2}{p}+p\right)=\tilde{P}_1\,,
P_2=0\,,P^3\equiv P=p-4\pi \frac{GM}{b}p= p+\tilde{P}\,.
\end{equation}
We calculate the power radiated as photons in the process of Fig.\ref{fig:Feynman2} according
to the formula \cite{renton}
\begin{equation}\label{W}
W=\frac{1}{8(2\pi)^2}\int \delta^{4}(P-p'-\ell)\frac{|M|^2}{Pp'_0}d^3p' d^3\ell\,.
\end{equation}
There are two ways to calculate $|M_1|^2$. In the first one we replace $p_{\alpha}$ with $P_{\alpha}$ in the
field-free ($\gamma_{\mu\nu}=0$) expression given by $\Sigma |M_1|^2 = Z^2 e^2 [-4m^2(p'_{\alpha}p^{\alpha})+8(p_{\alpha}p^{\alpha})]$. The gravitational contribution to $M_1$
then appears in $\tilde{P}_{\mu}$ exclusively.
We also remove the terms $-32m^2
(p'_{\alpha}p^{\alpha})+64m^2$ that do not contain gravitational contributions
and therefore refer to the kinematically forbidden transition.
This yields, to ${\cal O}(\gamma_{\mu\nu})$, the expression
\begin{equation}\label{M1sq}
\Sigma |M_1|^2 =Z^2e^2\left[-4(p'_{\alpha}\tilde{P}^{\alpha})+8
(p_{\alpha}\tilde{P}^{\alpha})\right]\,.
\end{equation}

In a second, alternate approach, we calculate $|M_1|^2$ directly ($\gamma_{\mu\nu}\neq 0$) from (\ref{M1}).
By summing over final spins and averaging over
initial spins and polarizations, we obtain
\begin{equation}\label{TR}
\Sigma |M_{1}|^{2} = \frac{Z^2e^2}{2(2m)^2}Tr\left\{(\mathbf{p'}+
m)\gamma_{\beta}\left[(\mathbf{p}+\mathbf{\tilde{P}}+
m)\left((\mathbf{p}+m)^2
+(\mathbf{p}+m)\mathbf{\tilde{P}^{*}}+H(\mathbf{p}+m)\right)\right]
\gamma^{\beta}\right\}\,,
\end{equation}
where $ \mathbf{a}=\gamma^{\mu}a_{\mu}$ and
$H=p^0/p^3 \phi (\gamma^3 p^0 -\gamma^0 p^3)+(p^0/p^3 \gamma^1 p^0
-\gamma^1 p^3)4GMK_1(bq_z)$. On carrying out the traces of the Dirac matrices, the contribution from $H$ vanishes. A similar, but simpler calculation, gives
$|M_{2}|^2$.
By further eliminating from $|M|^2$ the terms that
refer to the kinematically forbidden transition, we
find
\begin{equation}\label{TR1}
\Sigma |M|^2 =Z^2e^2\left\{\left[-4(p'_{\alpha}\tilde{P}^{\alpha})+8
(p_{\alpha}\tilde{P}^{\alpha})\right]-8\pi^2\frac{2GM}{b}\left[-4(p'_{\alpha}p^{\alpha})+8m^2\right]\right\}\,.
\end{equation}
The first set of square brackets in (\ref{TR1}) represents the contribution of $M_1$ and coincides with (\ref{M1sq}).
This supports our claim that the generalized momentum $P_{\mu}$ introduced in (\ref{mom}) leads to the correct value of the transition probability by the substitution of
$p_{\mu}$ with $P_{\mu}$ in the field-free expression. The second set of square brackets represents the contribution of $M_2$.

The integration over $d^3p'$ in (\ref{W}) is performed by means of the identity $\int \frac{d^3p'}{2p'_0}=\int d^4p'\delta(p'^2 -m^2)$,
while that over $\theta$ can be carried out by writing the on-shell condition for $p'$
in the form
\begin{equation}\label{onshell}
\delta(2|\vec{P}||\vec{\ell}|\cos\theta-P^{\alpha}P_{\alpha}
+2P_0\ell_0 + m^2)\,.
\end{equation}
We find
\begin{equation}\label{W1}
W=\frac{2Z^2e^2}{p^2}\frac{GM}{b}\left\{m^2\ell_0^2 -\left[\frac{m^2\ell_0^2}{2}+\frac{\ell_0^3(p_0-p)}{3}\right]\right\}\,.
\end{equation}
The first term in (\ref{W1}) represents the contribution of $M_1$.
The radiation spectrum is given by
\begin{equation}\label{spec}
\frac{dW}{d\ell_0}=\frac{2Z^2e^2\ell_0}{p^2}\left(\frac{GM}{b}\right)\left[m^2
-\ell_0\left(p_0-p\right)\right]\,.
\end{equation}
For $ p> m$, (\ref{W1}) gives
\begin{equation}\label{W2}
W_{p>m}= Z^2 e^2 \frac{GM}{b}\frac{m^2\ell_0^2}{p^2}\,,
\end{equation}
while it yields
\begin{equation}\label{W3}
W_{p<m}=Z^2 e^2 \frac{GM}{b} \frac{m \ell_0^2}{p^2}\left(\frac{m}{2}-\frac{\ell_0}{3}\right)\,,
\end{equation}
for $p<m $.
Equation (\ref{onshell}) and the condition $-1\leq \cos\theta \leq 1$ require that for $p>m$
\begin{equation}\label{ph}
4\pi p\left(\frac{GM}{b}\right)\left(1 +\frac{m^2}{4p^2}\right) \leq \ell_0 \leq 4\pi p\left(\frac{GM}{b}\right)\left(\frac{4p^2}{m^2}+3\right)\,.
\end{equation}
It follows from (\ref{ph}) that the hardest photons are emitted in the forward direction with energy $\ell_{0}\sim 16\pi(GM/b)p^{3}/m^{2}$ and
power $W_{p>m}\sim (16\pi Ze)^2(GM/b)^3 p^4/m^2$ which takes its highest values in the neighborhood of a black hole and for high values of $p/m$.

For $p<m$ we obtain
\begin{equation}\label{ph1}
4\pi\left(\frac{GM}{b}\right)m\left(1-\frac{p}{m}\right)\leq \ell_0 \leq 4\pi \left(\frac{GM}{b}\right)m\left(1+\frac{p}{m}\right)\,,
\end{equation}
which reduces to the single value $\ell_0\sim 4\pi m(GM/b)$ when $p \ll m$. In this case the radiation is still in the forward direction and the power radiated
\begin{equation}\label{W4}
W_{p<m}\sim 8(\pi Ze)^{2}(\frac{GM}{b})^{3} \frac{m^4}{p^2}\,,
\end{equation}
diverges for small values of $p$ (infrared divergence). This divergence arises as a consequence of the finite
energy resolution $\Delta\epsilon$ of the outgoing fermion. The process, as calculated, is in fact indistinguishable from that in which gravitons with energy $\leq \Delta\epsilon$
are also emitted and from processes in which vertex corrections are present (virtual gravitons emitted and reabsorbed by the external lines of Fig.\ref{fig:Feynman2}).
When these additional diagrams are calculated all infrared divergences disappear \cite{weinberg}.
In the particular case at hand, $p$ in (\ref{W4}) is simply replaced by $ 4\pi GMp_{0}/b$, as requested by
the external field approximation.

The results (\ref{W1})-(\ref{W3}) ignore the back-reaction on the background spacetime. We show below that this is not always negligible and provide an example of how a very small disturbance
in the wave equation can grow rapidly and alter the background gravitational field.

Equation (\ref{Psi}) requires that $\Psi_{0}(x)$ be a solution of the field-free Dirac equation and, of course, of the equation $(\eta^{\mu\nu}\partial_{\mu}\partial_{\nu}+m^2)\Psi_{0}(x)=0$.
The approximation procedure still holds true, however, when $\Psi_{0}(x)$ satisfies more general equations \cite{caipap0,papb}. With the addition of a dissipation term, the equation
for $\Psi_{0}$ becomes
\begin{equation}\label{damp}
\left(\eta^{\mu\nu}\partial_{\mu}\partial_{\nu}+m^2 -2m\sigma \partial_{0}\right)\Psi_{0}=0\,,
\end{equation}
where we take $\sigma=\alpha|\langle\Psi_{0}|\hat{T}|\Psi_{0}\rangle|^2 =\alpha(\frac{m}{p_0}\frac{GM}{2b})^2$ \cite{pap1} and $\alpha$ is a dimensionless, arbitrary parameter, $0\leq\alpha \leq 1$, that reflects the coupling strength of the dissipation term. When we substitute $\Psi_{0}(x)=\exp(m\sigma x_{0})\phi_{0}(x)$ into (\ref{damp}), we obtain
\begin{equation}\label{damp1}
\left[\partial_{0}^{2}-\partial_{z}^{2}+m^{2}(1-\sigma^{2})\right]\phi_{0}(x)=0\,.
\end{equation}
An example of a problem with similar behavior is offered by a fluid heated from below. For small temperatures gradients the fluid conducts the heat,
but as the gradient increases conduction is not sufficient to lead the heat away and the fluid starts to convect.
In realistic problems the exponential growth of $\Psi_{0}$ does not continue indefinitely, but is restricted at times $x_{0}>\tau \equiv 1/m\sigma$ by nonlinearities or dispersive effects that may have been initially neglected.

The effect of the new solution $\Psi_{0}$ on $W$ can be found as
follows. We first neglect the change $m\rightarrow
m\sqrt{1-\sigma^2}$ in $W$ because in general $\sigma<1$. The
effect of the exponentially increasing term on (\ref{W}) then
amounts to the transformations
$\delta(P_{0}-p_{0}'-\ell_{0})\rightarrow \delta(-2i\sigma
+P_{0}-p_{0}' -\ell_{0})$ and $\frac{1}{p^2}\rightarrow
\frac{1}{\beta^2 (p_{0}-2i\sigma)^2}\simeq
\frac{p_{0}^{2}}{\beta^{2}[(p_{0}^{2}-4\sigma^2)^2
+4p_{0}^{2}\sigma^{2}]}$, where we have used the relation
$\beta=p/p_0$. $W$ has therefore a resonance at $p_{0}=2\sigma$ of
width $4\sigma^2$. Over times $x_{0}>\tau=(\alpha
m)^{-1}(m/p_{0})^{-2}(GM/2b)^{-2} GeV^{-1}$,
$\Psi=\hat{T}\Psi_{0}$ increases exponentially until the
compensating mechanisms mentioned above kick in. For a proton of
energy $p_0\sim 10\,GeV$ in the field of a canonical neutron star
$ \tau\sim 3.5 \times 10^{-21}\alpha^{-1}\,\, s$. Considerably
higher values of $\tau$ can, of course, be obtained for the
lighter fermions. As $\Psi$ grows, so does the energy momentum
tensor associated with it and the gravitational field it
generates, altering, in the process, the gravitational background.

Summarizing, external gravitational fields in radiative processes can be included in the calculation of a
transition probability by simply replacing the momentum $p_{\mu}$ of a particle with its generalized version $P_{\mu}$
in the corresponding expression for the zero-field process. The example given refers to fermions,
but the procedure can be extended to particles of different spins. An essential point is here that the dispersion relations
are altered by the external gravitational
field and can be calculated if the corresponding wave equations
can be solved to ${\cal O}(\gamma_{\mu\nu})$, or higher \cite{caipap,pap1,pap2,pap3,pap4}. The treatment of particle lines in Feynman
diagrams therefore necessitates care when external
gravitational fields are present. It follows, in particular, that kinematically forbidden processes
like that of Fig.\ref{fig:Feynman2} become physical and their transition probabilities can be determined.
The calculation of the gravitational contributions are greatly simplified and can be extended to higher order in $\gamma_{\mu\nu}$.
The applications are not confined to fields of
a Newtonian type, but extend to any gravitational fields. In this respect, the procedure presented goes beyond the results that apply to external
electromagnetic potentials \cite{jauch}, not only because the gravitational field has in general ten components rather than just four,
but also because time-independence is not required in (\ref{Psi})-(\ref{M2}).

The procedure also yields transition amplitudes and decay rates that, to leading order, are linear in $\gamma_{\mu\nu}$ and can therefore be considerably larger
than those normally studied in the literature. These results suggest that some particle decay processes in the neighborhood
of compact astrophysical objects, or in cosmology, need to be re-evaluated.

We have also shown that the back-reaction of the fermion on the gravitational background need not be negligible. In fact, the addition of a small dissipation term in (\ref{damp})
can drastically transform the physical problem over a characteristic time $\tau$ in two respects: i) The power radiated by the fermion acquires a resonant, narrow peak at $p_{0}\sim 2\sigma$ and, ii)
$\Psi$ grows exponentially and, in so doing, affects the background gravitational field via its associated energy-momentum tensor, until so allowed by the physical circumstances.

\end{document}